\documentclass[aps,prd,preprintnumbers,superscriptaddress,nofootinbib,onecolumn,notitlepage]{revtex4-1}
\usepackage{ulem}
\usepackage[dvipdfmx]{graphicx}
\usepackage{bm,latexsym,amsmath,amssymb,amsfonts}
\usepackage[colorlinks=true,linkcolor=magenta,citecolor=magenta]{hyperref}
\usepackage{color}
\usepackage[utf8]{inputenc}

\newcommand{\calI}{{\cal I}}
\newcommand{\calO}{{\cal O}}

\newcommand{\pa}{{\partial}}

\newcommand{\vep}{{\varepsilon}}

\newcommand{\wt}[1]{{\widetilde{#1}}}
\newcommand{\wh}[1]{{\widehat{#1}}}
\newcommand{\ma}[1]{{\mathrm{#1}}}

\begin{document}

\title{On metric transformations with a $U(1)$ gauge field}

\author{Antonio De Felice}
\email{antonio.defelice@yukawa.kyoto-u.ac.jp}
\affiliation{Center for Gravitational Physics, Yukawa Institute for Theoretical
Physics, Kyoto University, Kyoto 606-8502, Japan}

\author{Atsushi Naruko}
\email{naruko@yukawa.kyoto-u.ac.jp}
\affiliation{Center for Gravitational Physics, Yukawa Institute for Theoretical
Physics, Kyoto University, Kyoto 606-8502, Japan}

\date{\today}

\preprint{YITP-19-108}

\begin{abstract}
  We study metric transformations including not just the field
  strength tensor of a $U(1)$ gauge field, but also its dual tensor.
  We first consider an arbitrary symmetric matrix built up with these
  two tensors in the metric transformation.  It turns out the form of
  transformation reduces to a quite simple form on imposing the parity
  evenness of the transformed metric and by utilising the
  Cayley-Hamilton theorem as well as other useful identities.
  Interestingly, the same form for the transformation was recently
  argued in the process of seeking for a generic metric transformation
  but without the inclusion of the dual tensor.
\end{abstract}

\maketitle

\section{Introduction}
Metric transformation is an interesting topic well beyond a merely
mathematical manipulation.  For example, under a conformal
transformation, although two frames will be physically the same as
well as any observable, the way of understanding physics may be
quite different as studied e.g.\ in \cite{Deruelle:2010ht,Chiba:2013mha}.

Such transformations, provided that they satisfy some properties
(e.g.\ the existence of their inverse transformations), in general do not change the
theory but its representation. Indeed, some representations of the theory may
give a deeper understanding of the theory itself. Something similar
happens in the choice of coordinates for a given background. A change
of variables/coordinates does not change the physics, but it may help
understanding the physics which takes place on the theory/background
under investigation. Filkenstein coordinates or Kruskal-Szekeres
coordinates are well known examples of coordinates which are able to
clearly show the physics, although they are not strictly necessary in
order to describe a Black Hole solution (see e.g.\ \cite{Misner:1974qy}).

Recently a disformal transformation draw much attention in the field
of scalar and vector fields coupled with gravity
\cite{Bekenstein:1992pj}.  Though it is equivalent to a change of the
time-coordinate in the case of homogeneous and isotropic universe, the
meaning of the same transformation remains unclear on a general
background. Related with this, a new type of disformal transformation
was recently introduced in \cite{Gumrukcuoglu:2019ebp}. In that paper,
the authors have introduced the exterior derivative of a one-form
field which is endowed with a $U(1)$ gauge invariance. The exact
two-form so obtained corresponds to the field strength tensor of the
$U(1)$ gauge field. It is clear that we can consider another two-form
under the same $U(1)$ gauge invariance, namely its dual tensor. This
is what we will do in this paper.

In a recent work, \cite{Deffayet:2013tca}, the authors have shown that
the galileon-like terms cannot be introduced in a flat spacetime under
the assumption of $U(1)$ invariance, although in a curved space-time
the vector Horndeski term can be allowed \cite{Horndeski:1976gi}.
Once we abandon the $U(1)$ gauge invariance, one can investigate a
broad class of theories of massive vector field
\cite{Heisenberg:2014rta,Allys:2015sht,Jimenez:2016isa,Heisenberg:2016eld,Kimura:2016rzw,Domenech:2018vqj}.
Theoretical developments in the theories of massless/massive vector fields may have interesting impacts for cosmology, such as cosmic inflation \cite{Watanabe:2009ct}, 
the generation of primordial magnetic fields on large scales (see \cite{Durrer:2013pga} for a review), 
or dark energy (see also e.g.\cite{deFelice:2017paw,Nakamura:2018oyy} for some interesting
phenomenology). 
By taking into account the case of massive vector field, possibilities for other more general
transformations will open up.
However, this would change the theory we are considering here. We will
then try to investigate how parity breaking objects, such as
pseudotensors can be playing some role in these general metric
transformations.

This paper is organised as follows.  We start with the introduction of
the notion of $(n \,, m)$ tensor, pseudotensors, and dual tensors of
anti-symmetric $p$-forms (see also \cite{Deffayet:2016von,Heisenberg:2019akx,Takahashi:2019vax} for recent developments of $p$-form theories). 
Then we discuss metric transformations
including the field strength tensor of a $U(1)$ gauge field as well as
its dual tensor. We find that generic transformations reduce to a
quite simple form by utilising several useful equations.  Finally we
conclude the paper.

\section{Parity, metric and some properties}

Parity transformations form a subset of the general coordinate transformation
between two coordinate frames. In particular, given a four-dimensional
coordinate transformation, which can be written in general as
\begin{equation}
x'^{\mu}=x'^{\mu}(x^{\nu})\,,
\end{equation}
then we can find how the differential quantities (that is the
infinitesimal displacements for each coordinate) transform as
\begin{equation}
dx'^{\mu}=\frac{\partial x'^{\mu}}{\partial x^{\nu}}\,dx^{\nu}\equiv\Lambda^{\mu'}{}_{\nu}\,dx^{\nu}\,,
\end{equation}
where $dx^{\mu}$ form the components of a vector field. Some
coordinate transformations may lead to a negative Jacobian, that is
$J\equiv\det(\Lambda^{\mu}{}_{\nu'})<0$, where
$\Lambda^{\mu}{}_{\nu'}$ is the inverse transformation of
$\Lambda^{\mu'}{}_{\nu}$, so that
$\Lambda^{\mu'}{}_{\nu}\Lambda^{\nu}{}_{\rho'}=\delta^{\mu'}{}_{\rho'}$,
and
$\Lambda^{\alpha}{}_{\mu'}\Lambda^{\mu'}{}_{\beta}=\delta^{\alpha}{}_{\beta}$.
Other transformations may lead instead to a positive value for $J$. In
general, we will only consider here the coordinate transformations for
which their Jacobian does not vanish, so that for any such
transformation its inverse exists and it is unique. Still, for a
general coordinate transformation, we could have different quantities, as we
shall show explicitly later on, which transform differently depending
on the sign of $J$.

On the other hand, just to be concrete and to give a well known
example, let us consider the metric, which is defined as (0,2)
symmetric tensor, $\bm{g}$, which acts linearly on two vectors in
order to give a scalar, as in
$\bm{g}(dx^{\alpha}\,\bm{e}_{\alpha},dx^{\beta}\,\bm{e}_{\beta})=g_{\mu\nu}dx^{\alpha}dx^{\beta}\bm{\omega}^{\mu}(\bm{e}_{\alpha})\,\bm{\omega}^{\nu}(\bm{e}_{\beta})=g_{\mu\nu}dx^{\alpha}dx^{\beta}\,\delta^{\mu}{}_{\alpha}\delta^{\nu}{}_{\beta}=g_{\mu\nu}dx^{\mu}dx^{\nu}=ds^{2}$.
Being a (0,2) symmetric tensor, its components are required to satisfy
the following transformation property
\begin{equation}
g_{\mu'\nu'}=\Lambda^{\rho}{}_{\mu'}\Lambda^{\sigma}{}_{\nu'}\,g_{\rho\sigma}\,,\qquad g_{\rho\sigma}=g_{\sigma\rho}\,.
\end{equation}
This is enough to imply that $g_{\mu'\nu'}=g_{\nu'\mu'}$, so that
after a general coordinate transformation, it remains symmetric. Finally,
on a pseudo Riemannian manifold, we will further impose that $g\equiv\det(g_{\mu\nu})<0$.
Therefore the determinant of the metric is always negative. Under
a general coordinate transformation we will have
\begin{equation}
g'=\det(g_{\mu'\nu'})=J^{2}\,g<0\,,
\end{equation}
so that the sign of $g$ becomes an invariant for any transformation
for which $J\neq0$.

Given a set of Cartesian-like coordinates -- this possibility is
clear at least locally for a local inertial frame -- for a pure parity
transformation in four dimensions, we assume $t'=t$, $x'^{i}=-x^{i}$.
In this case we find that $\Lambda^{\rho}{}_{\mu'}$ becomes diagonal
and, in particular, that $J=-1$, and this result does not change
on considering for example spherical coordinates. This transformation
does not affect the metric, so that the metric then is an object which
is even under this transformation.

In the following, we are about to introduce pseudotensors which differ
from tensors because of an extra multiplicative factor ${\rm sign}(J)$
in the usual transformation rule for their components between two
different coordinate frames. This extra factor will imply that pseudotensor
components transform, under parity, with an opposite (in sign) rule
if compared to tensors.

Let us now consider the difference among tensors and pseudotensors in more detail. We first recall here the definition of a $(n,m)$ tensor as an object
whose components under a general coordinate transformation transform
as 
\begin{equation}
T^{\alpha'_{1}\dots\alpha'_{n}}{}_{\beta'_{1}\dots\beta'_{m}}=\Lambda^{\alpha'_{1}}{}_{\rho_{1}}\cdots\Lambda^{\alpha'_{n}}{}_{\rho_{n}}\Lambda^{\sigma{}_{1}}{}_{\beta'_{1}}\cdots\Lambda^{\sigma_{m}}{}_{\beta'_{m}}\,T^{\rho{}_{1}\dots\rho{}_{n}}{}_{\sigma_{1}\dots\sigma{}_{m}}\,,
\end{equation}
were 
\begin{eqnarray}
\Lambda^{\alpha'}{}_{\rho} & = & \frac{\partial x'^{\alpha}}{\partial x^{\rho}}\,,\\
\Lambda^{\sigma}{}_{\beta'} & = & \frac{\partial x^{\sigma}}{\partial x'^{\beta}}\,,\\
\Lambda^{\alpha'}{}_{\rho}\Lambda^{\rho}{}_{\beta'} & = & \delta^{\alpha'}{}_{\beta'}\,,\\
\Lambda^{\sigma}{}_{\beta'}\Lambda^{\beta'}{}_{\rho} & = & \delta^{\sigma}{}_{\rho}\,,\\
J & \equiv & \det\!\left(\Lambda^{\sigma}{}_{\beta'}\right).
\end{eqnarray}
In particular, as already mentioned above, the metric tensor transforms as 
\begin{equation}
g_{\mu'\nu'}=\Lambda^{\rho}{}_{\mu'}\Lambda^{\sigma}{}_{\nu'}\,g_{\rho\sigma}\,.
\end{equation}
A tensor density of integer weight $w$ is defined instead as the object whose
components change according to the rule: 
\begin{equation}
\mathcal{T}^{\alpha'_{1}\dots\alpha'_{n}}{}_{\beta'_{1}\dots\beta'_{m}}=J^{w}\,\Lambda^{\alpha'_{1}}{}_{\rho_{1}}\cdots\Lambda^{\alpha'_{n}}{}_{\rho_{n}}\Lambda^{\sigma{}_{1}}{}_{\beta'_{1}}\cdots\Lambda^{\sigma_{m}}{}_{\beta'_{m}}\,\mathcal{T}^{\rho{}_{1}\dots\rho{}_{n}}{}_{\sigma_{1}\dots\sigma{}_{m}}\,,
\end{equation}
so that, for example, $g$ is a scalar density of weight 2.
For example out of the totally antisymmetric quantity $\epsilon_{\alpha\beta\gamma\mu}$ (with $\epsilon_{0123}=1$),
we can define
\begin{eqnarray}
\epsilon_{\alpha\beta\gamma\delta}\bm{d}x^{\alpha}\otimes\bm{d}x^{\beta}\otimes\bm{d}x^{\gamma}\otimes\bm{d}x^{\delta} & = & \epsilon_{\alpha\beta\gamma\delta}\Lambda^{\alpha}{}_{\rho'}\Lambda^{\beta}{}_{\sigma'}\Lambda^{\gamma}{}_{\mu'}\Lambda^{\delta}{}_{\nu'}\bm{d}x^{\rho'}\otimes\bm{d}x^{\sigma'}\otimes\bm{d}x^{\mu'}\otimes\bm{d}x^{\nu'}\nonumber \\
 & = & \epsilon_{\rho'\sigma'\mu'\nu'}J\bm{d}x^{\rho'}\otimes\bm{d}x^{\sigma'}\otimes\bm{d}x^{\mu'}\otimes\bm{d}x^{\nu'}\,,
\end{eqnarray}
so, by comparing the rhs of the last two lines, we find that the components
$\epsilon_{\alpha\beta\gamma\mu}$ transform as a (0,4) tensor density
of weight -1. Here we have used as basis for the covectors the 1-forms related to the coordinates, i.e.\ $\bm{\omega}^\alpha=\bm{d}x^\alpha$.

We can also define a pseudotensor as the object whose components
change according to: 
\begin{equation}
T^{\alpha'_{1}\dots\alpha'_{n}}{}_{\beta'_{1}\dots\beta'_{m}}={\rm sign}(J)\,\Lambda^{\alpha'_{1}}{}_{\rho_{1}}\cdots\Lambda^{\alpha'_{n}}{}_{\rho_{n}}\Lambda^{\sigma{}_{1}}{}_{\beta'_{1}}\cdots\Lambda^{\sigma_{m}}{}_{\beta'_{m}}\,T^{\rho{}_{1}\dots\rho{}_{n}}{}_{\sigma_{1}\dots\sigma{}_{m}}\,.
\end{equation}
Then if we define 
\begin{equation}
E_{\alpha\beta\gamma\delta}\equiv\sqrt{-g}\,\epsilon_{\alpha\beta\gamma\delta}\,,
\end{equation}
then, since $g'=J^{2}\,g$, we have 
\begin{eqnarray}
E_{\alpha\beta\gamma\delta}\bm{d}x^{\alpha}\otimes\bm{d}x^{\beta}\otimes\bm{d}x^{\gamma}\otimes\bm{d}x^{\delta} & = & \sqrt{-g}\epsilon_{\alpha\beta\gamma\delta}\Lambda^{\alpha}{}_{\rho'}\Lambda^{\beta}{}_{\sigma'}\Lambda^{\gamma}{}_{\mu'}\Lambda^{\delta}{}_{\nu'}\bm{d}x^{\rho'}\otimes\bm{d}x^{\sigma'}\otimes\bm{d}x^{\mu'}\otimes\bm{d}x^{\nu'}\nonumber \\
 & = & \frac{\sqrt{-g'}}{|J|}\epsilon_{\rho'\sigma'\mu'\nu'}J\bm{d}x^{\rho'}\otimes\bm{d}x^{\sigma'}\otimes\bm{d}x^{\mu'}\otimes\bm{d}x^{\nu'}\nonumber \\
 & = & \frac{1}{{\rm sign}(J)}E_{\rho'\sigma'\mu'\nu'}\bm{d}x^{\rho'}\otimes\bm{d}x^{\sigma'}\otimes\bm{d}x^{\mu'}\otimes\bm{d}x^{\nu'}\,,
\end{eqnarray}
which implies that $E_{\alpha\beta\gamma\delta}$ form the components
of a (0,4) pseudotensor. Out of this tensor, one can define the dual of a (2,0) antisymmetric tensor $M^{\mu\nu}$ as
\begin{equation}
  M_{\mu\nu}^{*}=\frac12\,E_{\mu\nu\alpha\beta}\,M^{\alpha\beta}\,.
\end{equation}

One can also define 
\begin{equation}
E^{\alpha\beta\gamma\delta}=\frac{1}{\sqrt{-g}}\,\bar\epsilon^{\alpha\beta\gamma\delta}\,,
\end{equation}
with $\bar\epsilon^{0123}=-\epsilon_{0123}=-1$, and show that $E^{\alpha\beta\gamma\delta}$ form the components of a $(4,0)$ pseudotensor\footnote{Here we have introduced the totally antisymmetric tensor density $\bar\epsilon^{\alpha\beta\gamma\delta}$ (of weight 1) which possesses useful properties if contracted with the other tensor density $\epsilon_{\alpha\beta\gamma\delta}$ introduced before (see also e.g.\ \cite{Misner:1974qy}). One can also show that $\bar\epsilon^{\mu\nu\rho\sigma}=-g g^{\mu\alpha} g^{\nu\beta} g^{\rho\gamma} g^{\sigma\delta} \epsilon_{\alpha\beta\gamma\delta}$.}. Out of this pseudotensor we can build up other pseudotensors such as 
\begin{equation}
 M^*{}^{\mu\nu}=\frac{1}{2}\,E^{\mu\nu\rho\sigma}\,M_{\rho\sigma}\,,
\end{equation}
 where $M_{\rho\sigma}$ are the components of an arbitrary antisymmetric (0,2) tensor, a 2-form.
Let us calculate 
\begin{eqnarray}
M^*_{\mu\alpha}\,M^*{}^{\alpha\nu} & = & \frac{1}{4}\,E_{\mu\alpha\rho\sigma}\,M^{\rho\sigma}\,E^{\alpha\nu\beta\gamma}\,M_{\beta\gamma}\nonumber \\
 & = & -\frac{1}{4}\,\epsilon_{\alpha\mu\rho\sigma}\,\bar\epsilon^{\alpha\nu\beta\gamma}\,M_{\beta\gamma}\,M^{\rho\sigma}\nonumber \\
 & = & \frac{1}{4}\,\delta_{\mu\rho\sigma}^{\nu\beta\gamma}\,M_{\beta\gamma}\,M^{\rho\sigma}
=\frac{1}{2}\,[M^{2}]\,\delta_{\mu}{}^{\nu}+M_{\mu\alpha}\,M^{\alpha\nu}\,.\label{dF2F2}
\end{eqnarray}
Here we have introduced a notation, $[M^{2}]\equiv M_{\mu\nu}M^{\mu\nu}$.
This leads to
\begin{eqnarray}
M^*_{\mu\alpha}\,M^*{}^{\alpha\beta}\,M^*_{\beta\gamma}\,M^*{}^{\gamma\nu} & = & \left(\frac{1}{2}\,[M^{2}]\,\delta_{\mu}{}^{\beta}+M_{\mu\alpha}\,M^{\alpha\beta}\right)\left(\frac{1}{2}\,[M^{2}]\,\delta_{\beta}{}^{\nu}+M_{\beta\gamma}\,M^{\gamma\nu}\right)\nonumber \\
 & = & \frac{1}{4}\,[M^{2}]^{2}\,\delta_{\mu}{}^{\nu}+[M^{2}]M_{\mu\alpha}\,M^{\alpha\nu}+M_{\mu\alpha}\,M^{\alpha\beta}\,M_{\beta\gamma}\,M^{\gamma\nu}\,.\label{dF4F4}
\end{eqnarray}
As a corollary of Eq.\ (\ref{dF2F2}),  we also find 
\begin{equation}
M^*_{\mu\nu}\,M^*{}^{\mu\nu}=-M_{\rho\sigma}\,M^{\rho\sigma}=-[M^{2}]\,,\label{trdF2F2}
\end{equation}
and 
\begin{equation}
M^*_{\mu\alpha}\,M^*{}^{\alpha\beta}\,M^*_{\beta\gamma}\,M^*{}^{\gamma\mu}=M_{\mu\alpha}\,M^{\alpha\beta}\,M_{\beta\gamma}\,M^{\gamma\mu}\equiv [M^{4}]\,.\label{trdF4F4}
\end{equation}
It is not difficult to show that $M^*_{\mu\alpha}\,M^{\alpha\nu}$
is proportional to identity. This is only true if $M$ is antisymmetric.
For example 
\begin{eqnarray}
M^*_{0\alpha}\,M^{\alpha1} & = & \frac{1}{2}\,E_{0\alpha\rho\sigma}\,M^{\rho\sigma}M^{\alpha1}=\frac{1}{2}\,E_{0j\rho\sigma}\,M^{\rho\sigma}M^{j1}\nonumber \\
 & = & \frac{1}{2}\,E_{02\rho\sigma}\,M^{\rho\sigma}M^{21}+\frac{1}{2}\,E_{03\rho\sigma}\,M^{\rho\sigma}M^{31}\nonumber \\
 & = & \frac{1}{2}\,(E_{0213}\,M^{13}+E_{0231}\,M^{31})M^{21}+E_{0312}\,M^{12}M^{31}\nonumber \\
 & = & E_{0213}\,M^{13}M^{21}+E_{0312}\,M^{21}M^{13}=0\,.
\end{eqnarray}
Along the same lines we can also show that the diagonal elements are
all equal to each other, and that 
\begin{align}
M_{\mu\rho}M^{*}{}^{\rho\nu}=-\frac{1}{4}(M_{\alpha\beta}M^{*}{}^{\alpha\beta})\delta_{\mu}{}^{\nu}=-\frac{1}{4}[MM^{*}]\delta_{\mu}{}^{\nu}\,,\label{FdF}
\end{align}
As a corollary we also find 
\begin{equation}
M_{\mu}{}^{\alpha}\,M_{\alpha}{}^{\beta}\,M_{\beta}{}^{\nu}=M_{\mu}{}^{\alpha}\,M_{\alpha\beta}\,M^{\beta\nu}=M_{\mu}{}^{\alpha}\,\left(M^*_{\alpha\beta}\,M^*{}^{\beta\nu}-\frac{1}{2}[M^{2}]\,\delta_{\alpha}{}^{\nu}\right)=-\frac{1}{4}[MM^{*}]\,M^{*}{}_{\mu}{}^{\nu}-\frac{1}{2}[M^{2}]\,M_{\mu}{}^{\nu}\,.\label{F3}
\end{equation}
We are now ready to consider the metric transformations we want to discuss.

\section{metric transformation with a dual tensor}

The field strength of a $U(1)$ gauge field will be defined in terms of
a vector potential $A_{\mu}$ as
\begin{align}
F_{\mu\nu}\equiv\pa_{\mu}A_{\nu}-\pa_{\nu}A_{\mu}\,.
\end{align}
Its dual tensor can be defined by 
\begin{align}
F^{*}{}^{\mu\nu}\equiv\frac{1}{2\sqrt{-g}}\bar{\vep}^{\mu\nu\rho\sigma}F_{\rho\sigma}\,.
\end{align}
It is interesting to note that 
\begin{align}
F_{\mu\rho}F^{*}{}^{\rho\nu}=-\frac{1}{\sqrt{-g}}(E_{i}B^{i})\delta_{\mu}{}^{\nu}=-\frac{1}{4}(F_{\alpha\beta}F^{*}{}^{\alpha\beta})\delta_{\mu}{}^{\nu}\,,
\end{align}
which is one of key equations in this paper. Here we have introduced
\begin{align}
E_{i}\equiv F_{0i}\,,\qquad B^{i}\vep_{ijk}=-F_{jk}\,,
\end{align}
where $\vep_{ijk}$ is the totally antisymmetric tensor in 3D and
$\vep_{123}=1$. Then we find that
\[
[FF^{*}]=F_{\mu\rho}F^{*}{}^{\mu\rho}=\frac{4}{\sqrt{-g}}\,E_{i}B^{i}\,.
\]
At the same time we also have
\begin{eqnarray}
\ma{det}[F{}_{\mu\nu}] & = & (E_{i}B^{i})^{2}\,,\\
\ma{det}[F^{\mu}{}_{\nu}] & = & \frac{1}{g}\,\ma{det}[F{}_{\mu\nu}]=\frac{(E_{i}B^{i})^{2}}{g}=-\frac{1}{16}\,[FF^{*}]^{2}\,.
\end{eqnarray}

Another interesting equation which is nothing but the Cayley-Hamilton
theorem in $4$D is 
\begin{align}
F^{\mu}{}_{\rho}F^{\rho}{}_{\sigma}F^{\sigma}{}_{\delta}F^{\delta}{}_{\nu}+\frac{1}{2}(F_{\alpha\beta}F^{\alpha\beta})\,F^{\mu}{}_{\rho}F^{\rho}{}_{\nu}+\ma{det}[F^{\mu}{}_{\nu}]\,\calI^{\mu}{}_{\nu}=\calO^{\mu}{}_{\nu}\,,\label{CH}
\end{align}
where $\calI^{\mu}{}_{\nu}$ and $\calO^{\mu}{}_{\nu}$ represent
the identity and zero matrices respectively. By taking the trace of
this equation, one gets 
\begin{align}
[F^{4}]=\frac{1}{2}[F^{2}]^{2}-4\,\ma{det}[F^{\mu}{}_{\nu}]=\frac{1}{2}[F^{2}]^{2}+\frac{1}{4}[FF^{*}]^{2}\,.\label{F4}
\end{align}

Let us consider a generic metric transformation constructed by the
field strength tensor of a $U(1)$ gauge field and its dual. First,
any matrix composed by $F^{\mu}{}_{\nu}$ can be simply represented
by a superposition of $\calI^{\mu}{}_{\nu}\,,F^{\mu}{}_{\nu}\,,(F^{2})^{\mu}{}_{\nu}$
and $(F^{3})^{\mu}{}_{\nu}$ thanks to the above theorem. In fact,
any higher order matrix will reduces to this form: 
\begin{align}
\overbrace{F^{\mu}{}_{\rho_{1}}\cdots F^{\rho_{n-1}}{}_{\nu}}^{n}=c_{3}\,(F^{3})^{\mu}{}_{\nu}+c_{2}\,(F^{2})^{\mu}{}_{\nu}+c_{1}\,F^{\mu}{}_{\nu}+c_{0}\,\calI^{\mu}{}_{\nu}\,,
\end{align}
where $c_{0\,,\cdots\,,3}$ will be functions of $[F^{2}]$ and $[F^{4}]$.
Needless to say, also as for its dual tensor, any matrix composed
by $F^{*}{}^{\mu}{}_{\nu}$ will reduce to a similar form: 
\begin{align}
\overbrace{F^{*}{}^{\mu}{}_{\rho_{1}}\cdots F^{*}{}^{\rho_{n-1}}{}_{\nu}}^{n}=d_{3}\,(F^{*}{}^{3})^{\mu}{}_{\nu}+d_{2}\,(F^{*}{}^{2})^{\mu}{}_{\nu}+d_{1}\,F^{*}{}^{\mu}{}_{\nu}+d_{0}\,\calI^{\mu}{}_{\nu}\,,
\end{align}
where $d_{0\,,\cdots\,,3}$ will be functions of $[F^{*}{}^{2}]$
and $[F^{*}{}^{4}]$. However thanks to \eqref{trdF2F2} and \eqref{trdF4F4},
we can identify $d_{0\,,\cdots\,,3}$ as functions of $[F^{2}]$ and
$[F^{4}]$.

Next, let us consider an arbitrary matrix which can be defined by
$F^{\mu}{}_{\nu}$ and $F^{*}{}^{\mu}{}_{\nu}$: 
\begin{align}
 & \overbrace{F^{\mu}{}_{\rho_{1}}\cdots F^{\rho_{n_{1}-1}}{}_{\mu_{2}}}^{n_{1}}\overbrace{F^{*}{}^{\mu_{2}}{}_{\sigma_{1}}\cdots F^{*}{}^{\sigma_{m_{1}-1}}{}_{\mu_{3}}}^{m_{1}}\cdots\overbrace{F^{\mu_4}{}_{\alpha_{1}}\cdots F^{\alpha_{n_{s}-1}}{}_{\mu_5}}^{n_{s}}\overbrace{F^{*}{}^{\mu_5}{}_{\beta_{1}}\cdots F^{*}{}^{\beta_{m_{s}-1}}{}_{\nu}}^{m_{s}}\,.
\end{align}
By utilising the Cayley-Hamilton theorem, \eqref{CH}, it is not hard
to imagine that the above tensor reduces to this form 
\begin{align}
 & \sum_{N_{1}\,,\cdots N_{s}\,,M_{1}\,,\cdots M_{s}}\overbrace{F^{\mu_{1}}{}_{\rho_{1}}\cdots F^{\rho_{N_{1}-1}}{}_{\mu_{2}}}^{N_{1}\leq3}\overbrace{F^{*}{}^{\mu_{2}}{}_{\sigma_{1}}\cdots F^{*}{}^{\sigma_{M_{1}-1}}{}_{\mu_{3}}}^{M_{1}\leq3}\cdots\overbrace{F^{\mu_4}{}_{\alpha_{1}}\cdots F^{\alpha_{N_{s}-1}}{}_{\mu_5}}^{N_{s}\leq3}\overbrace{F^{*}{}^{\mu_5}{}_{\beta_{1}}\cdots F^{*}{}^{\beta_{M_{s}-1}}{}_{\nu}}^{M_{s}\leq3}\,.
\end{align}
Then with the aid of \eqref{FdF}, it will further reduce to the superposition
of $\calI^{\mu}{}_{\nu}$,$F^{\mu}{}_{\nu}$, $(F^{2})^{\mu}{}_{\nu}$
and $(F^{3})^{\mu}{}_{\nu}$ and also $F^{*}{}^{\mu}{}_{\nu}$,$(F^{*}{}^{2})^{\mu}{}_{\nu}$
and $(F^{*}{}^{3})^{\mu}{}_{\nu}$: 
\begin{align}
 & e_{3}\,(F^{3})^{\mu}{}_{\nu}+e_{2}\,(F^{2})^{\mu}{}_{\nu}+e_{1}\,F^{\mu}{}_{\nu}+f_{3}\,(F^{*}{}^{3})^{\mu}{}_{\nu}+f_{2}\,(F^{*}{}^{2})^{\mu}{}_{\nu}+f_{1}\,(F^{*})^{\mu}{}_{\nu}+f_{0}\,\calI^{\mu}{}_{\nu}\,,
\end{align}
since $F^{\mu}{}_{\rho}F^{*}{}^{\rho}{}_{\nu}\propto\delta^{\mu}{}_{\nu}$.
Here $e$ and $f$ are functions of $[F^{2}]$, $[F^{4}]$, $[F^{*}{}^{2}]$,
$[F^{*}{}^{4}]$ and $[FF^{*}]$. Finally by utilising \eqref{dF2F2}
and \eqref{F3} to eliminate $(F^{3})^{\mu}{}_{\nu}$, $(F^{*})^{\mu}{}_{\nu}$
and $(F^{*}{}^{2})^{\mu}{}_{\nu}$, one gets 
\begin{align}
 & \overbrace{F^{\mu_{1}}{}_{\rho_{1}}\cdots F^{\rho_{n_{1}-1}}{}_{\mu_{2}}}^{n_{1}}\overbrace{F^{*}{}^{\mu_{2}}{}_{\sigma_{1}}\cdots F^{*}{}^{\sigma_{m_{1}-1}}{}_{\mu_{3}}}^{m_{1}}\cdots\overbrace{F^{\mu_4}{}_{\alpha_{1}}\cdots F^{\alpha_{n_{s}-1}}{}_{\mu_5}}^{n_{s}}\overbrace{F^{*}{}^{\mu_5}{}_{\beta_{1}}\cdots F^{*}{}^{\beta_{m_{s}-1}}{}_{\nu}}^{m_{s}}\nonumber \\
 & =\wh{e}_{2}\,(F^{2})^{\mu}{}_{\nu}+\wh{e}_{1}\,F^{\mu}{}_{\nu}+\wh{f}_{1}\,(F^{*})^{\mu}{}_{\nu}+\wh{f}_{0}\,\calI^{\mu}{}_{\nu}\,.\label{great}
\end{align}
Here coefficients can be regarded as functions of $[F^{2}]$, $[F^{4}]$
and $[FF^{*}]$ only since $[F^{*}{}^{2}]$ and $[F^{*}{}^{4}]$ can
be replaced by $[F^{2}]$ and $[F^{4}]$ by \eqref{trdF2F2} and \eqref{trdF4F4}.

Now it is the time to consider a metric transformation including $F$
as well as $F^{*}$ tensors. Since a metric tensor is a symmetric
and parity even quantity, the generic form of a metric transformation
will be given by 
\begin{align}
\wt{g}_{\mu\nu} & =\omega \, g_{\mu\nu}+\gamma_{1} \, F_{\mu\rho}g^{\rho\sigma}F_{\sigma\nu}+\gamma_{2} \, F_{\mu\rho}^{*}g^{\rho\sigma}F_{\sigma\nu}^{*}+\frac{1}{2}\gamma_{3} \, (F_{\mu\rho}g^{\rho\sigma}F_{\sigma\nu}^{*}+F_{\nu\rho}g^{\rho\sigma}F_{\sigma\mu}^{*})\times[FF^{*}]\,,
\end{align}
where $\omega$ will be defined by 
\begin{align}
\omega=\omega\Bigl([F^{2}]\,,[F^{4}]\,,[FF^{*}]^{2}\Bigr)\,,
\end{align}
and $\gamma_{1\,,2\,,3}$ are similarly defined. These scalar functions have no explicit
dependence on $[F^{*}{}^{2}]$ and $[F^{*}{}^{4}]$ by \eqref{trdF2F2}
and \eqref{trdF4F4}. First, it should be noted that a liner term
in $F_{\mu\nu}$ or $F_{\mu\nu}^{*}$ cannot be introduced due to
the antisymmetric nature of indices. Any higher order matrices of
$F$ and $F^{*}$ will reduce to a matrix at most quadratic in $F$
as shown in \eqref{great}. Moreover $F_{\mu\rho}^{*}g^{\rho\sigma}F_{\sigma\nu}^{*}$
can be replaced by $F_{\mu\rho}g^{\rho\sigma}F_{\sigma\nu}$ using
\eqref{dF2F2} and $F_{\mu\rho}g^{\rho\sigma}F_{\sigma\nu}^{*}$ is
proportional to $g_{\mu\nu}$. Finally, thanks to \eqref{F4}, the
dependence of coefficients on $[FF^{*}]^{2}$ can be eliminated. Thus
we conclude that the most generic metric transformation build up by
$F$ and $F^{*}$ can be simply expressed by 
\begin{align}
\wt{g}_{\mu\nu} & =\Omega \bigl([F^{2}]\,,[F^{4}]\bigr) \, g_{\mu\nu}+\Gamma\bigl([F^{2}]\,,[F^{4}]\bigr) \, F_{\mu\rho}g^{\rho\sigma}F_{\sigma\nu}\,.\label{result}
\end{align}
Before the end of this section, we note about the invertibility of thus obtained transformation.
Since the form of transformation is the same as that in \cite{Gumrukcuoglu:2019ebp}, the invertibility condition will be also the same as the one found there.

\section{summary}

In this short note, we have discussed metric transformations including not just the field strength tensor of a $U(1)$ gauge field, $F_{\mu \nu}$, but also its dual tensor, $F^*_{\mu \nu}$. 
At first we consider an arbitrary symmetric matrix built up with $F_{\mu \nu}$ and $F_{\mu \nu}^*$ in the metric transformation.
Thanks to the Cayley-Hamilton theorem and several equations related with the dual tensor,
 for example $F_{\mu \rho} g^{\rho \sigma} F_{\sigma \nu}^* \propto g_{\mu \nu}$, it turned out that the form, $\wt{g}_{\mu\nu} = \Omega \, g_{\mu\nu} + \Gamma \, F_{\mu\rho}g^{\rho\sigma}F_{\sigma\nu}$ shown in \eqref{result} where $\Omega$ and $\Gamma$ are functions of $[F^2]$ and $[F^4]$,
  is the most generic transformation of metric constructed by $F_{\mu \nu}$ and $F_{\mu \nu}^*$. Interestingly, the same form was recently argued in seeking for a generic metric transformation but only focusing on $F_{\mu \nu}$. Here we have verified that the form discussed there can be the most generic one even if we take into account the dual tensor of the field strength tensor of a U(1) gauge.
Any application to the early universe cosmology or mysterious dark energy is left for future study.

\acknowledgments 
A.N.\ would like to thank Chunshan Lin for the kind invitation to the conference
"Beyond General Relativity, Beyond Cosmological Standard Model"
held in Warsaw where this work was initiated. The work of A.N.\ was
partly supported by JSPS KAKENHI Grant No.\ JP19H01891.




%

\end{document}